\documentclass[aps,showpacs,showkeys,twocolumn]{revtex4}
\usepackage{graphicx}
\usepackage{amssymb}
\usepackage{bm}

\begin{document}

%%%%%%%%%%%%%%%%%%%%%%%%%%%%%%%%%%%%%%%%%%%%%%%%
\newcommand{\bear}{\begin{eqnarray}}
\newcommand{\eear}{\end{eqnarray}}
%%%%%%%%%%%%%%%%%%%%%%%%%%%%%%%%%%%%%%%%%%%%%%%%

\setcounter{page}{1}
\title[]{Spacetime structure of 5D hypercylindrical vacuum solutions with tension}

\author{Inyong \surname{Cho}}
\email{iycho@skku.edu} \affiliation{BK21 Physics Research Division
and Department of Physics, Sungkyunkwan University, Suwon 440-746,
Korea}

\author{Gungwon \surname{Kang}}
\email{gwkang@kisti.re.kr}
\affiliation{Korea Institute of Science and Technology Information,
52-11 Eoeun-dong, Yuseong-gu, Daejeon 305-806, Korea}

\author{Sang Pyo \surname{Kim}}
\email{sangkim@kunsan.ac.kr}
\affiliation{Department of Physics,
Kunsan National University, Kunsan
573-701, Korea}

\author{Chul H. \surname{Lee}}
\email{chulhoon@hanyang.ac.kr}
\affiliation{Department of Physics and BK21 Division of Advanced Research
and Education in Physics, Hanyang University, Seoul 133-791, Korea}

\date[]{}

\begin{abstract}
We investigate geometrical properties of 5D cylindrical vacuum
solutions with a transverse spherical symmetry. The metric is
uniform along the fifth direction and characterized by tension and
mass densities. The solutions are classified by the
tension-to-mass ratio. One particular example is the well-known
Schwarzschild black string which has a curvature singularity
enclosed by a horizon. We focus mainly on geometry of other
solutions which possess a naked singularity. The light signal
emitted by an object approaching the singularity reaches a distant
observer with finite time, but is infinitely red-shifted.
\end{abstract}



\maketitle

\section{Introduction}
Black string solutions are higher dimensional
black hole spacetimes possessing
``hypercylindrical" horizons with/without compactification,
instead of ``spherical" ones.
Recently, it has been of much interest studying properties
of those spacetime backgrounds.
Being different from that the stationary black hole
with a spherical horizon topology is stable,
the Schwarzschild black string
was found to be unstable under small perturbations;
it is the so-called Gregory-Laflamme (GL)
instability~\cite{Gregory:1993vy}. This instability has been studied
in many ways afterwards. In particular,
whether a perturbed black string is fragmented into an array of
small black holes, or ends up with a stable non-uniform black string
has been a hot issue~\cite{Horowitz:2001cz,BSBH}.
The robustness of the GL instability has also been studied in
supergravity theories~\cite{Hirayama:2002hn} as well as
in general relativity with a negative cosmological
constant~\cite{Hirayama:2001bi}.
However, it is still not understood well what really causes the GL
instability.

The Schwarzschild black string is a particular case
of the 5D hypercylindrical vacuum solution.
It is characterized by a single parameter (usually called $M$)
while the general solution has two parameters.
The two-parameter solution was first
found by Kramer~\cite{Kramer:1971ik}
and was manipulated in various ways by
others in the literature~\cite{Chodos:1980df,
Gross:1983hb,Davidson:1985zf,Yoshimura:1986dd,Lee:2006jx}.
Although geometrical propeties of this spacetime
were studied in many works, most of the
studies were in the context of the Kaluza-Klein thoery.
Consequently, understanding of the geometry was based on the
four-dimensional gravity with a scalar field.
This caused many misleading interpretations
for the full five-dimensional geometry of the solutions.

Very recently, the physical meaning of the two parameters
was correctly interpreted for the first time
in Ref.~\cite{Lee:2006jx} by one of us.
The author considered the weak-field solutions
of the Einstein field equations outside
some matter distribution.
He identified the two parameters with
``mass" and ``tension" densities by matching with
the vacuum solutions at asymptotic region.
When the tension-to-mass ratio is exactly one half,
in particular, it corresponds to the Schwarzschild black string.

In this paper we investigate geometrical properties of such
``hypercylindrical" spacetime having arbitrary tension in detail.
The solutions are classified mainly by the tension-to-mass ratio $a$.
The physical range of the ratio is $0\leq a \leq 2$.
Specific values $a=1/2$ and $a=2$ correspond to the well-known
Schwarzschild black string and Kaluza-Klein bubble~\cite{Elvang:2004iz}.
We are particularly interested in the other values of $a$ in the range.
The geometry possesses a naked singularity.
The light signal emitted by an observer approaching the singularity
escapes within finite time, but is infinitely redshifted.
There is no wormhole structure in the full five dimensional geometry.

In Section II, we study the geometrical properties
of the hypercylindrical solution.
In Section III, we discuss the causal structure of the spacetime,
and we conclude in Section IV.

\section{Geometrical Properties}
The most general form of the  static metric
for the transverse spherically symmetric static spacetime
with a translational symmetry along the fifth spatial direction
in five dimensions may be written as
\bear
ds^2 = -Fdt^2 +G\left[ d\rho^2 +\rho^2 \left( d\theta^2
+\sin^2\theta d\phi^2 \right) \right] +Hdz^2 .
\label{metric}
\eear
Here $F$, $G$ and $H$ are functions of
the ``isotropic" radial coordinate $\rho$ only.
Note that the fifth direction
is not assumed to be flat in general, i.e., $H \ne 1$.
If we include a constant momentum flow along the z direction,
the $g_{tz}$ component is not zero in general.
Such a stationary solution was considered
in Refs.~\cite{Chodos:1980df, Lee:2007wu}.
Time-dependent solutions in a separable form were
found in Ref.~\cite{Liu:1993xg}.
A class of solutions allowing the $z$-dependence
was also considered in Ref.~\cite{Billyard:1996dj}.
By taking double Wick rotations, i.e., $t \to iz$ and $z \to it$
in the metric~(\ref{metric}), one can easily
see that, given a solution with $F$ and $H$,
the metric with $F$ and $H$ being exchanged is a
solution as well.

The hypercylindrical type of system has been
studied by a number of people,
and its vacuum solution has been obtained in various forms.
The solutions are basically two-parameter solutions.
The interpretation of these two integration constants
has not been given properly for a long time.
In Ref.~\cite{Chodos:1980df} these constants were related
to the gravitational mass and scalar charge.
Davidson and Owen~\cite{Davidson:1985zf}
defined two kinds of mass parameters in
the context of Kaluza-Klein dimensional reduction.
Namely, they defined the gravitational
mass parameter by considering the asymptotic behavior
of the four-dimensional effective metric,
and speculated that the other mass parameter
is somehow related to the Kaluza-Klein electric charge.

It is Ref.~\cite{Lee:2006jx} in which the physical meaning of
these integration constants was correctly given.
Using the metric ansatz (\ref{metric}),
the solutions are given by
\bear
F(\rho) &=& \left|
\frac{1-\frac{K}{\rho}}{1+\frac{K}{\rho}}
\right|^{\frac{2(2 -a)}{\sqrt{3(a^2 -a +1)}}},
\label{F}\\
G(\rho) &=& \left(
1+\frac{K}{\rho} \right)^4 \left|
\frac{1+\frac{K}{\rho}}{1-\frac{K}{\rho}} \right|^{\frac{2(a+1)}{\sqrt{3(a^2
-a +1)}} -2},
\label{G}\\
H(\rho) &=& \left|
\frac{1-\frac{K}{\rho}}{1+\frac{K}{\rho}} \right|^{\frac{2(2a
-1)}{\sqrt{3(a^2 -a +1)}}}.
\label{H}
\eear
In Ref.~\cite{Lee:2006jx}, the author considered weak-field nonvacuum solutions
of five dimensional Einstein equations for a system of matter
distribution characterized by the mass and tension densities
having the same spherical and translational symmetries mentioned above.
By comparing them with the asymptotic behaviors of the metric
components above at spatial infinity ($\rho \gg K$),
he found the relationship between these two integration constants
and the physical parameters of the linear-mass density $\lambda$
and the linear-tension density $\tau$,
\bear
a &=& {\tau \over \lambda},\\
K &=& \sqrt{1-a+a^2 \over 3} G_5\lambda .
\eear

Actually, this identification is a sort of analogy
in the sense that the internal-vacuum region
is replaced by a compact matter having the same symmetries.
Therefore, the mass and tension in this analogy
are contributions from the matter stress-energy inside,
but are not pure gravitational contributions.
The rigorous definitions of mass and tension densities for gravitational fields
themselves can be found in Refs.~\cite{Traschen:2001pb, Townsend:2001rg, Harmark:2004ch}
where the ADM-tension density is associated with
the asymptotic spatial-translation symmetry
along the $z$ direction in much the same way as
that the ADM-mass density is associated with
the asymptotic time-translation symmetry.
Such definitions give the same relationship above.

The geometrical properties of the spacetime
under consideration are very different
depending on the value of the tension-to-mass ratio $a$.
($K$ can be absorbed to the radial coordinate $\rho$.)
The particular case of $a = 1/2$ stands for
the well-known Schwarzschild black string,
\bear
ds^2 = &-&\left( {1-K/\rho \over 1+K/\rho} \right)^2dt^2 \nonumber\\
{} &+& \left( 1+{K\over \rho}\right)^4 (d\rho^2 +\rho^2 d\Omega_2^2) +dz^2, \\
= &-&\left( 1- {2M\over r}\right) dt^2 \nonumber\\
{} &+& \left( 1- {2M\over r}\right)^{-1}dr^2
+r^2d\Omega_2^2 +dz^2,
\label{Schmetric}
\eear
where $M = 2K = G_5\lambda$ and $r = \rho (1+K/\rho)^2$
is the usual ``isentropic" circumference radius.
The case of $a = 2$ stands for
the so-called Kaluza-Klein bubble~\cite{Elvang:2004iz}.

A priori there is no restriction on the parameter values.
However, if we assume the positivity of the mass density,
we have $\lambda \geq 0$.
In Ref.~\cite{Traschen:2003jm} it was proved that the pure gravitational contribution
to the tension is positive definite.
Therefore, the physical range for the values of the tension
would be $\tau \geq 0$ (i.e., $a \geq 0$).
From the weak-field approximation studied in Ref.~\cite{Lee:2006jx},
in a relation to the matter contribution to the tension,
one can show that there exists an upper bound
in the tension provided that the matter satisfies
the strong-energy condition.
This upper bound gives $\tau \leq 2\lambda$ (i.e., $a \leq 2$)
in five dimensions.
We apply the same upper bound for the gravitational contribution
to the tension although there is no definite proof for
it in the literature as long as we know.
Therefore, one may assume that the physical range
of the tension parameter is
\bear
0 \leq \tau \leq 2\lambda \quad
(\mbox{i.e., }0 \leq a \leq 2).
\eear

Based on some desirable cosmological behavior, Davidson and
Owen~\cite{Davidson:1985zf} speculated that the physical choice is
$a < 1/2$. On the other hand, Ponce de
Leon~\cite{PoncedeLeon:2006xs} claimed that $-1 < a < 1/2$ by
imposing the physical energy conditions on the four-dimensional
effective matter induced in the Kaluza-Klein dimensional reduction.

Regarding the geometrical properties, first note that the
Kretschmann scalar $R_{ABCD}R^{ABCD}$ diverges at $\rho =K$
except for $a=1/2$ and $a=2$.
Therefore, a curvature singularity locates there.

\begin{figure}
\begin{center}
\includegraphics[height=55mm]{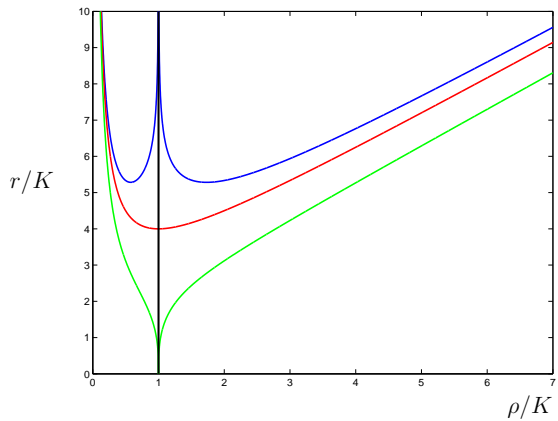}
\end{center}
\caption{Plot of $r$ vs. $\rho$.
From the top down, the curves show the typical shape corresponding
to the case of
$1/2 < a <2$, $a=1/2, 2$, and $0<a<1/2$.
}
\label{fig1}
\end{figure}

Now let us  consider the isentropic radius $r$
defined by
\bear
r = \rho \sqrt{G(\rho)} = \frac{(\rho +K)^2}{\rho}\left|
\frac{\rho+K}{\rho-K} \right|^{\frac{(a+1)}{\sqrt{3(a^2 -a +1)}} -1}.
\label{r}
\eear
The shape of $r(\rho)$ is classified into three types
depending on the scale of $a$
as shown in Fig.~\ref{fig1}.
%%%%%%%%%%%%%%%%%%%%%%%%%%%%%%%%%%%%%%%%%%%%%%%%%%%%%%%%%%%%%%%%%%%%%%
%%%%%%%%%%%%%%%%%%%%%%%%%%%%%%%%%%%%%%%%%%%%%%%%%%%%%%%%%%%%%%%%%%%%%%
As we can see from the figure, there exist two copies of the spacetime
separated by $\rho=K$. Both copies share the same geometry
since the metric is invariant under the transformation $\rho \to K/\rho^2$.
Therefore, we shall focus on the region of $\rho \geq K$ from now.

For the Schwarzschild black string ($a=1/2$),
the $\rho = K$ surface corresponds to the event horizon at $r=2M$ $(=4K)$
from Eq.~(\ref{Schmetric}).
The $\rho$-coordinate cannot describe the interior region of the horizon
while the $r$-coordinate can do.
The Kaluza-Klein bubble $(a=2)$ solution is related to the
Schwarzschild black string simply by double Wick rotations.

For $0<a <1/2$, the space for $r=[0,\infty)$ is well described
by $\rho =[K,\infty)$, and those two radial coordinates
have a one-to-one correspondence.

For $1/2 <a <2$, the relation between $r$ and $\rho$ is very peculiar.
For a given value of $r$, $\rho$ is double valued.
There exists an extremum at
\bear
\rho_+ = \frac{a+1 + \sqrt{(2a-1)(2-a)}}{\sqrt{3(a^2-a+1)}} K.
\label{rho+}
\eear
The region below $r(\rho_+)$ is not covered by the $\rho$-coordinate.
The $S^2$ surface area at $z = {\rm constant}$
hypersurface is $A(\rho) = 4\pi r^2$.
For $1/2 <a <2$, this area is infinite at $\rho =K$, reaches minimum at
$\rho=\rho_+$, and then increases again.
Therefore, it looks like a wormhole geometry at $\rho_+$
and attracted people's attention in earlier works.

The proper length along the fifth direction is given by
\bear
L(\rho) = \int^{z_0+1}_{z_0} \sqrt{H}dz = \left|
\frac{1-\frac{K}{\rho}}{1+\frac{K}{\rho}} \right|^{\frac{2a -1}{\sqrt{3(a^2
-a +1)}}} .
\eear
As $\rho$ decreases, it monotonically shrinks
down to zero at $\rho =K$ for $a > 1/2$ whereas it expands to
infinity for $a < 1/2$.

The total surface area of a cylindrical sector having a coordinate
distance $\Delta z = 1$ is then given by
\bear
A_{\rm total}(\rho) &=& A(\rho)\times L(\rho) = 4\pi r^2 \sqrt{H}
\nonumber\\
{} &=& 4\pi \frac{(\rho
+K)^4}{\rho^2} \left| \frac{1-\frac{K}{\rho}}{1+\frac{K}{\rho}}
\right|^{2-{3 \over \sqrt{3(a^2 -a +1)}}} .
\eear
Since the exponent above is always positive except for $a=1/2$,
the total area turns out to be zero at $\rho=K$ for $a \not= 1/2$,
and increases monotonically as $\rho$ increases.
For $a=1/2$, the exponent becomes zero and the surface area is finite.
Therefore, although the submanifold at $z={\rm constant}$ looks like
a wormhole geometry as explained above, the full geometry including
the $z$ direction is not that of a wormhole spacetime.

\section{Causal structure}
In this section we discuss the causal structure
of the given geometry.
The interesting things would occur at $\rho=K$ where
the singularity is located, and at $\rho_+$ where
the $S^2$ surface area becomes minimum.

Let us consider radial motions at a $z={\rm constant}$ submanifold.
The metric becomes
\bear
ds^2 = -Fdt^2 +Gd\rho^2
= -F\left( dt +d\rho^{\ast}\right)\left( dt -d\rho^{\ast} \right),
\label{guv}
\eear
where the tortoise coordinate $\rho^{\ast}$ is defined as
\bear
d\rho^{\ast} = \sqrt{\frac{G}{F}}d\rho .
 \label{tortoise}
\eear
The ingoing- and outgoing-null coordinates are defined respectively as
\bear
v = t +\rho^{\ast} \quad {\rm and} \quad u=t -\rho^{\ast}.
\label{nullcoord}
\eear
In the vicinity of $\rho=K$ we have
\bear
\sqrt{\frac{G}{F}} \sim \left| \rho -K \right|^q ,
\eear
where
\bear
q=1 -{\sqrt{3}\over\sqrt{a^2-a+1}}
\Rightarrow
\left\{%
\begin{array}{ll}
    q=1 & (a=1/2), \\
    -1<q<1 & (a \neq 1/2). \\
\end{array}%
\right.
\eear
Therefore, the tortoise coordinate becomes
\bear
\rho^\ast &\sim& \ln |\rho-K| \quad\;\; (a=1/2), \\
{} &\sim& |\rho-K|^{{q+1}} \quad (a \neq1/2).
\eear
For the Schwarzschild black string ($a=1/2$),
$\rho^{\ast}$ goes to negative infinity as $\rho \to K$.
Consequently, from Eq.~(\ref{nullcoord}),
the ingoing-null geodesic ($v ={\rm constant}$)
touches the $\rho = K$ surface at $t = \infty$,
indicating an event horizon there.

\begin{figure}
\begin{center}
\includegraphics[height=55mm]{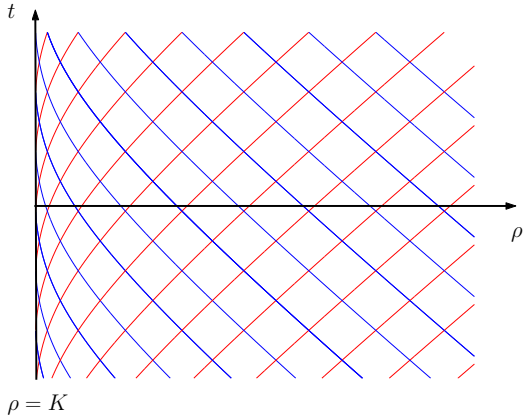}
\end{center}
\caption{Plot of ingoing- and outgoing-null geodesics for $-1 < a <
2$ ($a\neq 1/2$). The null rays touch $\rho =K$ at a finite time.
The geodesics were numerically integrated.} \label{fig2}
\end{figure}

For $a \not= 1/2$, however,
$\rho^\ast$ is finite as $\rho \to K$.
It implies that the value of $t$ along the ingoing-null geodesic
becomes finite on reaching $\rho =K$.
This indicates that the coordinate $t$ is
not singular there as can be seen in Fig.~\ref{fig2}.
(Note that the proper time measured by a fixed observer
outside is proportional to $t$.)
%%%%%%%%%%%%%%%%%%%%%%%%%%%%%%%%%%%%%%%%%%%%%%%%%%%%%%%%%%%%%%%%%%%%%%
%%%%%%%%%%%%%%%%%%%%%%%%%%%%%%%%%%%%%%%%%%%%%%%%%%%%%%%%%%%%%%%%%%%%%%

\begin{figure}
\begin{center}
\includegraphics[height=45mm]{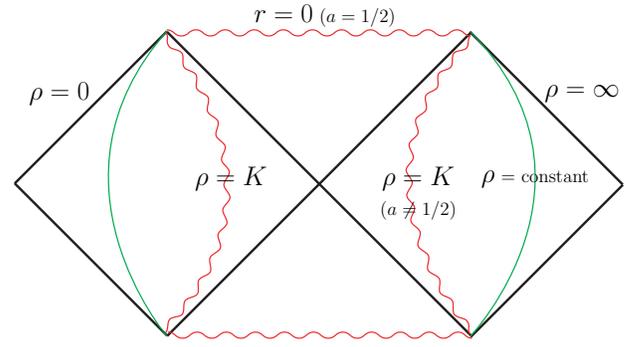}
\end{center}
\caption{Penrose diagram at a $z = {\rm constant}$ slice.}
\label{fig3}
\end{figure}

In order to see whether or not the null rays
can escape from the $\rho =K$ surface,
let us now consider the behavior of an outgoing-null
geodesic in $(v, \rho)$ coordinates,
\bear
ds^2 = -Fdt^2 +Gd\rho^2
= -Fdv\left( dv -2\sqrt{\frac{G}{F}}d\rho\right).
\label{gvrho}
\eear
Integrating the outgoing-null geodesic
$du = dv -2\sqrt{G/F}d\rho = 0$
from $\rho =K +\epsilon$, the value of $v$
(a well-defined time coordinate)
elapsed for reaching $\rho = \rho_0$ becomes
\bear
\Delta v &=& \int dv = 2\int^{\rho_0}_{K+\epsilon} \sqrt{\frac{G}{F}}
d\rho \\
&\sim & \left\{
\begin{array}{ccc}
\ln \left|\frac{\rho_0 -K}{\epsilon}\right| \qquad\qquad\;\;\,   (a=1/2),  \\
\left|\rho_0 -K\right|^{q+1} -\left|\epsilon\right|^{q+1} \;(a \neq 1/2).
\end{array}
\right. \eear Note that $\Delta v \to \infty$ as $\epsilon \to 0$
for the case of $a=1/2$. It implies the existence of an event
horizon at $\rho =K$, which is well-known for the case of the
Schwarzschild black string. On the other hand, $\Delta v$ is finite
as $\epsilon \to 0$ for the case of $a \not= 1/2$. It implies that
the light signal can actually escape from the $\rho = K$ surface
within fine time as it was discussed earlier. Therefore, $\rho = K$
is not an event horizon, but a naked singularity in this family of
spacetime solutions. The global causal structure at a $z={\rm
constant}$ slice is illustrated in its Penrose diagram in
Fig.~\ref{fig3}.

Although the light signal takes a finite time to escape from
the singularity, the lapse function in the metric~(\ref{gvrho})
vanishes there, which indicates it is an infinite-redshift surface.
Therefore, the light signal emitted from the singularity gets
infinitely redshifted~\cite{footnote1}.

The next question is whether or not the singularity at $\rho =K$ is
naked. Usually one easy way to search an event horizon in a
spherically symmetric system is to look up where the $g_{rr}$
component diverges in isentropic coordinates. According to the
coordinate transformation~(\ref{r}), the spherically symmetric
metric of the 3D sector becomes \bear ds_3^2 = G(d\rho^2 +\rho^2
d\Omega_2^2) ={dr^2 \over \left( 1+ {\rho \over 2G}{dG \over d\rho}
\right)^2} +r^2 d\Omega_2^2 . \eear After some algebra, we can show
that the place where the $g_{rr}$ component diverges corresponds to
$\rho =\rho_+$ defined in Eq.~(\ref{rho+}). However, it is easy to
see from Eq.~(\ref{guv}) that this $\rho_+$ surface is not a null
surface. Therefore, $\rho = \rho_+$ is not an event
horizon~\cite{footnote2}.
It is not very difficult to see that there exists no other candidate
for the event horizon in this geometry. Therefore, the $\rho = K$
surface is a naked singularity.

%%%%%%%%%%%%%%%%%%%%%%%%%%%%%%%%%%%%%%%%%%%%%%%%%%%%%%%%%%%%%%%%%%%%%%
%%%%%%%%%%%%%%%%%%%%%%%%%%%%%%%%%%%%%%%%%%%%%%%%%%%%%%%%%%%%%%%%%%%%%%

\section{Conclusions}
In this paper, we studied a five dimensional
cylindrical system in vacuum.
We assumed the translational symmetry along the fifth direction
and the spherical symmetry along
transverse 4D sectors.
The solutions to the Einstein equations of such a system have two
parameters which correspond to mass and tension densities. The
solutions are classified by the tension-to-mass ratio
$a=\tau/\lambda$. We concluded that the physically meaningful range
is $0 \leq a \leq 2$ while particular values correspond  to the
Schwarzschild black string ($a=1/2$) and the Kaluza-Klein bubble
($a=2$).

Adopting a metric ansatz in isotropic coordinates for the 3D sector,
there exist two identical copies of the spacetime.
We investigated interesting geometrical features
compared with those of the Schwarzschild black string.
Except for $a = 1/2$ and $2$,
there exists a curvature singularity at $\rho =K$.
From the singularity,
the total surface area of a cylindrical sector gradually increases
from zero as $\rho \rightarrow \infty$ or $\rho \rightarrow 0$.
It means that the previously-known wormhole structure in considering
the transverse $S^2$ surface, does not appear in the full geometry.
In fact, such a wormhole structure is a consequence of the
dimensional reduction to the 4D Jordan frame. If we transform it to
the 4D Einstein frame, the wormhole structure is naturally absent.
We also point out that even in the 4D Jordan frame the proper length
in the radial direction between the throat ($\rho = \rho_+$)
and $\rho = K$ is not infinite, but is finite.

While the $\rho =K$ surface of the Schwarzschild black string
is an event horizon, it is not true for the others.
The outgoing-null rays emitted from the singularity take
finite time to escape.
In addition, there is no event horizon in the outer geometry either.
Therefore, the singularity at $\rho =K$ is naked.
However, the lapse function becomes zero at the singularity,
so the outgoing-null rays get infinitely redshifted.

Another interesting issue related with this hypercylindrical system
would be investigating its stability.
Although it is known that the Schwarzschild black string
is unstable to perturbations and experiences the Gregory-Laflamme instability,
the physical cause of the instability has not been revealed yet.
While the Schwarzschild black string is described by only one parameter,
the general hypercylindrical solution that we have
now has two parameters.
The complete knowledge of the role of the parameters,
particularly the tension density, may help in understanding
the physical cause of the black string instability.
We will come back to this issue in future work.

\section*{Acknowledgments}
This work was supported
by the Basic Research Program
of the Korea Science \& Engineering Foundation
under the grant No.~R01-2006-000-10965-0 and
Astrophysical Research Center for the Structure and
Evolution of the Cosmos (ARCSEC) (I.C.),
in part by Grand Challenge Project at KISTI (G.K.),
by the Korea Science and Engineering Foundation (KOSEF) grant funded
by the Korea government (MOST)
under the grant No. R01-2005-000-10404-0 (S.P.K.), and
No. R01-2006-000-10651-0 (C.H.L.),
and in part by the Asia Pacific Center for Theoretical Physics.

\end{document}